\documentclass[twocolumn,showpacs,prl,superscriptaddress]{revtex4}
\usepackage{graphicx}
\usepackage{bm}

\def\bea{\begin{eqnarray}}
\def\eea{\end{eqnarray}}

\def\nn{\nonumber}

\def\la{\langle}
\def\ra{\rangle}

\def\om{\omega}

\def\g{\gamma}

\def\f{\frac}

\def\mT{{\mathcal{T}}}

\begin{document}

\title{Energy current magnification in coupled oscillator loops}
\author{Rahul Marathe}
\email{rahul.marathe@mpikg.mpg.de}
\affiliation{Departamento de Fisica Atomica, Molecular y Nuclear and GISC, Universidad Complutense de Madrid, 28040
-Madrid, Spain}
\author{Abhishek Dhar}
\email{dabhi@rri.res.in}
\affiliation{Raman Research Institute, Bangalore 560080, India}
\author{A. M. Jayannavar}
\email{jayan@iopb.res.in}
\affiliation{Institute of Physics, Bhubaneshwar 751005, India}
\date{\today} 
\begin{abstract}
Motivated by studies on current magnification  in quantum mesoscopic 
systems we consider sound and heat transmission in classical models of
oscillator chains. A loop of coupled oscillators is connected to two leads
through which one can either transmit monochromatic waves or white noise
signal from heat baths. We look for the    
possibility of current magnification in this system due to some asymmetry
introduced between the two arms in the loop. We find that current
magnification is indeed obtained for particular frequency ranges. However the
integrated current  shows the effect only in the presence of a pinning
potential for the atoms in the leads. We also study the effect of
anharmonicity on current magnification.
\end{abstract}
\maketitle

Mesoscopic systems of micron size have recently been studied extensively.
In these systems, at low temperatures the mean free path of an electron can 
exceed the sample dimensions, thus maintaining the coherence of the single 
particle wave function across the entire sample. In such coherent
systems several novel effects have been observed which do not have any
classical analogue \cite{WW92,Im02}. An interesting example is that of a
mesoscopic conducting 
loop connected through metallic leads to two separate electron reservoirs at
different chemical potentials $\mu_L$ and $\mu_R$ respectively. A current is
established between these two reservoirs. In the presence of this transport
current it has been shown that current magnification (CM) can occur 
in the loop \cite{Jayan1PRB:95,JayanPRB:95}. In this case under appropriate
conditions depending on the  Fermi energy, the current in one arm of the
loop exceeds the total current across the system.
For current conservation at the 
junctions, the current in the other arm flows in the reverse direction,
{\emph {i.e}}  opposite  to the transport current in the leads. 
The predicted circulating current density can be very large \cite{JayanPRB:95}
and has been
termed as giant persistent current \cite{YWHL01}. 
Several studies have shown the existence of CM in  a
variety of models including multi-channel systems \cite{JayanPRB:04}, and
systems with   spin  \cite{JayanIJMPB:98,ZGL05} and  thermoelectric
\cite{moskalets98} currents.  
\begin{figure}
\vspace{0.5cm}
\includegraphics[width=3.3in]{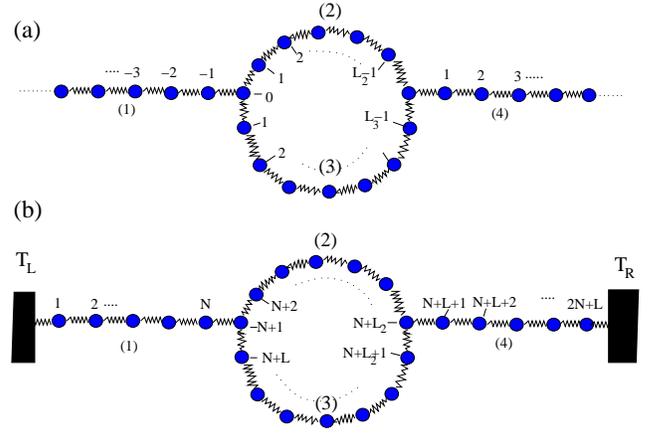}
\caption{(color online). Models studied: (a) Oscillator loop connected to
  infinite leads formed by coupled oscillators. (b) System with finite leads
  connected to heat reservoirs. } 
\label{sys}
\end{figure}

Motivated by these studies on quantum systems, in this paper we address the
question as to whether CM can occur in classical models of
energy transport in oscillator networks. Energy transport in such systems is
by lattice vibrations which in the long wavelength limit corresponds to
sound-waves. Since CM is basically a wave-phenomena,
one expects that it should be observable also in oscillator networks.  
Heat conduction in such networks have earlier been studied by
Eckmann and Zabey \cite{EckmannJSP:04} who also discussed the possibility  of
circulating  currents. 
The system we consider is schematically shown in Fig.~(\ref{sys}). It consists
of a loop formed by particles, all with  masses $m$ and connected by springs
with stiffness $k$ . Two Leads [$(1)$ and $(4)$ in the
figure] formed by semi-infinite oscillator chains are connected to the loop in
an asymmetric way such that the loop has two arms $(2)$ and $(3)$ of unequal
lengths. We consider two situations. First we study the transmission of single
frequency plane waves across this geometry. Next we consider the case where a
band of frequencies are fed into the leads by connecting them to heat baths
kept at different temperatures. We also look at the effect of anharmonicity
on CM. Since CM is a wave-phenomena one expects it to be suppressed in
the presence of any thermalizing  
mechanism such as arising from inelastic phonon-phonon scattering due to
anharmoncity.
In fact for macroscopic systems these kind of processes would lead to
an Ohmic (diffusive) behavior of the wire in which case one would not get
CM. In the 
electron case one important source of 
thermalization is electron-electron interactions. However it is quite difficult
to study its effect on CM without using approximation methods. On the
other hand for the classical oscillator case, it is easy to introduce
anharmonicity and numerically see its effect on CM.

\begin{figure}
\includegraphics[width=3.3in]{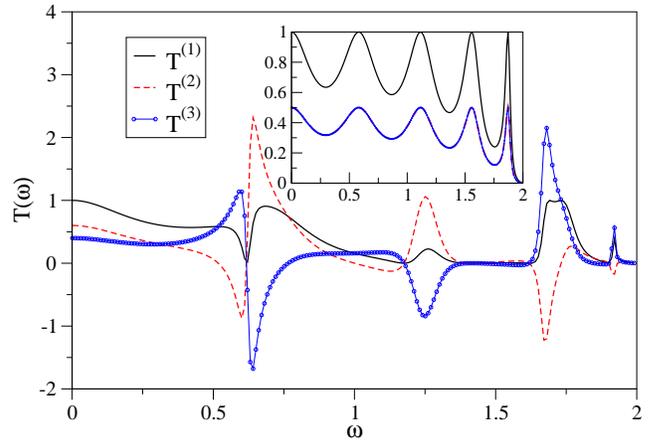}
\caption{(color online). Plot of transmission-versus-frequency  in different
  regions of the network for the case of infinite leads. 
Parameters used are $k=1$, $m=1$, $L_2=4$ and $L_3=6$. Inset shows the same
plots for the  case where there is no asymmetry in the loop arm lengths 
( $L_2=L_3=6$).} 
\label{TvswSound}
\end{figure}

We first discuss transmission of  monochromatic plane waves 
across the system. The number of particles on the loop are $L=L_2+L_3$ and
particles in different regions are numbered as in 
Fig.~(\ref{sys}a). For a wave incident from the left side  the particle
displacements are given by:
\bea
x^{(1)}_{l} &=& Re[~(e^{iql_1}~+~r e^{-iql_1})~e^{-i\om \tau}~] ~,~ -\infty
< l_1 \leq 0\nn\\ 
x^{(2)}_{l} &=& Re[~(t_1~e^{iql_2} ~+~ r_1~ e^{-iql_2})~e^{-i\om \tau}~] ~,~ 0 \leq l_2 \leq L_2\nn\\
x^{(3)}_{l} &=& Re[~(t_2~e^{iql_3} ~+~ r_2~ e^{-iql_3})~e^{-i\om \tau}~] ~,~
 0 \leq l_3 \leq L_3\nn\\
x^{(4)}_{l} &=& Re[~t~e^{iql_4}~e^{-i\om \tau}]  ~,~ 0 \leq l_4 < \infty,\label{dispsol}
\eea       
with $ q \in (0,\pi)$ and $\om = 2 \sqrt\f{k}{m}\ \sin(q/2)$, the usual
dispersion relation for the harmonic chain. 
The six unknown transmission amplitudes  can be
found out by matching the solutions Eqns.~(\ref{dispsol}) 
at the junctions and considering the equations of motion of the  
particles at the junctions.
Then one gets:
\begin{widetext}
\bea
1 ~+~ r &=& t_1~ +~ r_1 =  t_2 ~+~ r_2 \nn\\
-m~w^2~(~1~ +~ r~) &=& -k~(~3(1+r)-~e^{-iq}-~r~ e^{iq}-~t_1~ 
e^{iq}-~r_1~ e^{-iq}-~t_2~ e^{iq}-~r_2~ e^{-iq})\nn\\
t~ &=& ~t_1~ e^{iq~L_2}+~r_1~ e^{-iq~L_2} =  ~t_2~ e^{iq~L_3}+~r_2~ e^{-iq~L_3}\nn\\
-m~w^{2}~t &=& -k~( 3t ~- te^{iq}-~ t_1 e^{iq~(L_2-1)}-~r_1 e^{-iq~(L_2-1)}
-~ t_2 e^{iq~(L_3-1)}-~r_2 e^{-iq~(L_3-1)}).\nn\\
\eea
\end{widetext}
These linear equations can be explicitly solved  to get all the unknown 
amplitudes. 
 We do not 
give explicit expressions here since they are quite lengthy. 
The instantaneous energy current from site  $l$ to $l+1$ is the
product of the velocity of the  
$(l+1)^{\rm th}$ particle  and the force on it due to the $l^{\rm th}$
particle. 
Thus the energy current averaged over a time period between two neighboring
sites   on any of the four regions ($s=1,2,3,4$) is given by:
 $I^{(s)}= -k ~(\om/2\pi)\int_{0}^{2
  \pi/\om} ~d \tau~ \dot{x}_{l+1}^{(s)}~[~x_{l+1}^{(s)}-x_{l}^{(s)}~]$. 
One then finds  $I^{(s)} 
= (k \omega \sin{q}/2) \mT^{(s)}(\omega)$ where the transmission coefficients
are given by $\mT^{(1)}=\mT^{(4)}=1-|r|^2=|t|^2$, $\mT^{(2)}= |t_2|^2-|r_2|^2$ and
$\mT^{(3)}= |t_3|^2-|r_3|^2$~.  
In Fig.~(\ref{TvswSound}) we plot these transmission coefficients as a function of
frequency $\om$ for a particular choice of parameters $k, m, L_2, L_3$ with  
$L_2 < L_3$. The most interesting features that we see are that, for certain
values of the frequency,  the transmission $\mT^{(2)}$ [$\mT^{(3)}$]  
on one of the arms  of the loop can be {\emph{negative}} and when this happens
there is CM on the other arm, {\emph{i.e}} $|\mT^{(3)}/\mT^{(1)}| > 1 $
[$|\mT^{(2)}/\mT^{(1)}| > 1$]. 
In contrast, the inset of Fig.~(\ref{TvswSound}) gives results for $L_2=L_3$ in
which case all transmission coefficients are positive with magnitudes less
than one. Note that current conservation implies
$I^{(1)}=I^{(2)}+I^{(3)}=I^{(4)}$ and hence negative current flow on one arm
necessarily implies CM on the other.  

Next we ask the question as to what happens when the two leads are connected to
white-noise heat baths at different temperatures $T_L$ and $T_R$. Do we still
get negative heat current flow in one of the arms and CM ? 
For systems with diffusive heat flow this is clearly not possible. In the
nonequilibrium steady state one would expect the system to be in local thermal
equilibrium which 
means that the local temperatures at the junctions will define the direction
of heat flow and this will then be unique (high-to-low temperature). However,
for  ballistic heat flow, as in a harmonic system,  a local temperature does not
have a thermodynamic significance and there is a possibility that we can 
get   CM. We again consider the harmonic network shown in
Fig.~(\ref{sys}b) with each lead consisting of a finite number of
particles, say $N$. The full network of the loop and leads thus has
$M=2N+L$ particles and is described by the harmonic Hamiltonian 
\bea
\mathcal{H}&=& \frac{1}{2}\ \dot{X}^T {\bf M} \dot{X} + \frac{1}{2}\ X^T {\bm \Phi} X, \label{hamiltonian} 
\eea
where $X=\{x_1,x_2,...,x_M\}$,  ${\bf M}$ is the diagonal mass matrix and
${\bm \Phi}$ is the force 
matrix. We label the particles as shown in Fig.~(\ref{sys}b). 
To model the heat baths, the particles at the free ends of
the leads have extra terms in their equations of motion corresponding to a
Langevin dynamics. Thus the particle at the end of the left reservoir has an
extra part $-\gamma_L \dot{x}_1 +\eta_L$ in its equations of motion
while the particle at the right end has an extra part $-\gamma_R
\dot{x}_M +\eta_R$. The noise terms are Gaussian with zero mean and
variances given by the fluctuation-dissipation relations 
$\la\eta_{L,R}(t)\eta_{L,R}(t')\ra = 2k_BT_{L,R}\g_{L,R}\delta_{L,R}\delta(t-t')$. 
For the same parameter values as in Fig.~(\ref{TvswSound}) we 
computed the average heat current in the four regions. The steady state heat
current on a bond between sites $a$ and $b$ is given by $J^{(s)}=-k \la 
\dot{x}_{b}^{(s)} 
[~x_{b}^{(s)}-x_{a}^{(s)}~] \ra $ where $\la...\ra$ now denotes a noise
average.  Using the methods described in Ref~\cite{dharroy06}, the heat
current in any part of the system can be expressed as an integral  over all
frequencies of the transmission function, which can  
be written in terms of the Green's function matrix
$G(\omega)=[-\omega^2 
  {\bf M} + {\bm \Phi} -i \omega {\bm \Gamma}]^{-1} $ where ${\bm \Gamma}$ is a
dissipation matrix whose only non-vanishing elements are the two diagonal terms
$\gamma_L$ and $\gamma_R$ occurring at positions corresponding to the two sites
connected to reservoirs.        
Thus, assuming that a unique nonequilibrium steady-state is achieved, the
heat current on any part  can be written as \cite{dharroy06}: 
\bea
I^{(s)} = \f{k_B(T_L -T_R)}{4\pi}\ \int_{-\infty}^{\infty} d\om~
\mT^{(s)} (\om) ,\label{Ileads}
\eea 
where the transmission coefficient $\mT^{(s)}(\omega)=  2 k i \om ~
[~G_{a,1}(\om)G_{b,1}(-\om)   -   ~G_{a,M}(\om)G_{b,M}(-\om)~]$
with $a,b$ being two adjacent sites on the region of interest. The site $b$ is
chosen to be on the right of $a$ so that the convention used is that current
is from left-to-right.

\begin{figure}
\includegraphics[width=3.3in]{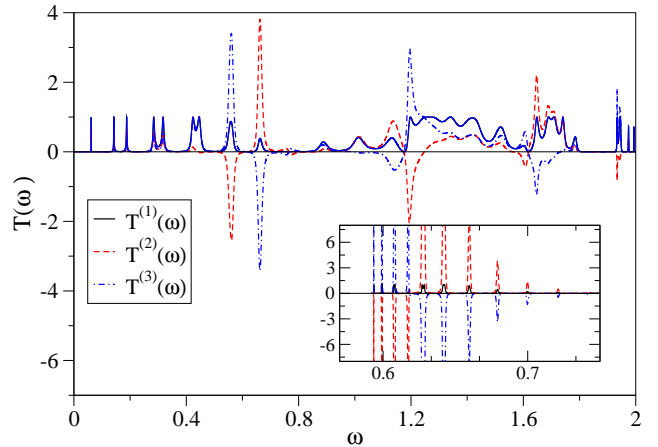}
\caption{(color online). Plot of transmission-versus-frequency in different
  regions   of the network for the same system  as in   Fig.~(\ref{TvswSound}) but
  with finite leads ($N=20$) 
  connected to Langevin heat reservoirs at temperatures $T_L=10$ and
  $T_R=1$. The inset   shows $\mT(\omega)$ for the case where the parameters of
  the leads are chosen  such that they have a narrow frequency transmission
  band (see text for the parameter values). } 
\label{TvswHB}
\end{figure}

In all the calculations we set with $T_L=10, T_R=1$ and $N=20$. The other
parameters of the network are the same as for the data in Fig.~(\ref{TvswSound}).  
In Fig. (\ref{TvswHB}) we plot the transmission coefficients in  the leads
and in the two arms. We see that for particular values of frequencies the
value of 
$\mT^{(2)}(\omega)$ is negative and whenever this occurs we have
$|\mT^{(3)}/\mT^{(1)}| > 1$ implying CM. 
Next we calculate the total heat current using Eqs.~(\ref{Ileads}). 
From our numerical calculations   we find $I^{(1)}= 0.7661, I^{(2)}= 0.37968, 
I^{(3)}=0.3864$. 
We have also verified these results from
direct nonequilibrium simulations. Thus we do not get any CM. 
 For several other choices of parameter values we find the same
situation. Introducing impurities in the loop can  
enhance the asymmetry in the system. With this we find that while one gets 
CM in specific frequency ranges, the integrated current again does not
show any CM.

\begin{figure}
\includegraphics[width=3.3in]{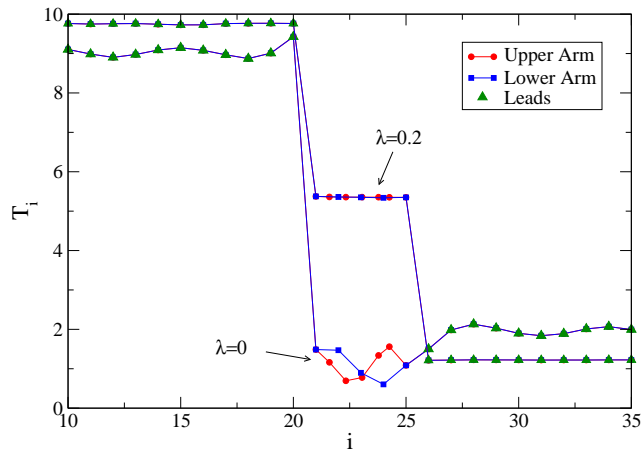}
\caption{(color online). Plot of the temperature profile in the network for
  the  same system  as considered in the inset of  Fig.~(\ref{TvswHB})
  without anharmonicity ($\lambda=0$) and with anharmoncity ($\lambda=0.2$).
The temperature across the leads and the two arms of the loop are shown. }
\label{Tprof}
\vspace{0.25cm}
\end{figure}

Let us now consider the effect of introducing a band-pass filter between the
loop and reservoirs such that only frequencies over the range where CM
occurs, are allowed to pass.   
It seems plausible that introduction of such a filter will allow us to observe
CM in the integrated current. We now test this
numerically. For this we introduce a harmonic pinning potential on all sites
of the leads. Thus for particles in the leads, in addition to the
interparticle potential $k(x_l-x_{l+1})^2/2$, there is also an additional
onsite potential $V(x_l)=k_ox_l^2/2$.  
For the choice of pinning strength $k_o=0.35$ and $k=0.1$  (only on the
leads), only frequencies in the 
range $\omega \approx (0.6-0.87)$ can pass through the leads. This is the
range where, as 
can be seen from the plot of $\mT(\omega)$ data for the unpinned system in
Fig.~(\ref{TvswHB}),  we expect maximum  CM.
In the inset of Fig.~(\ref{TvswHB}) we show plots of $\mT^{(1)},\mT^{(2)},\mT^{(3)}$
for 
this system. As expected, we find significant transmission in a small
frequency band. The values of the integrated currents in this case are:
 $I^{(1)}=0.01245, I^{(2)} = 0.06681, I^{(3)} =  -0.05436$.
Thus in this case CM is observed even for the integrated current.    
Note that there is CM on both the arms and the magnification factor is more
than about five times. 
In Fig.~(\ref{Tprof}) we show the temperature profile in this network with
pinning. Here we define a local temperature at any site through its mean
kinetic energy, {\emph{i.e}} $T_l=m \la \dot{x}_l^2 \ra$ at the $l^{\rm th}$ site.   
We see that the temperature profile is non-monotonic. As mentioned earlier
this is not surprising since the local temperature in this integrable system
does not have the usual thermodynamic meaning.

\begin{figure}
\vspace{0.5cm}
\includegraphics[width=3.3in]{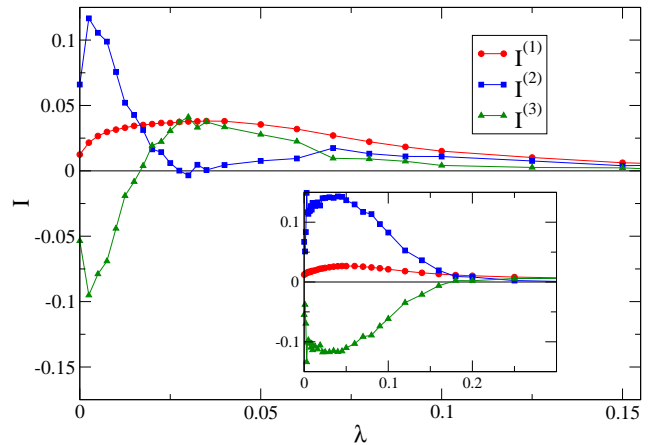}
\caption{(color online). Plots of currents $I^{(1)}, I^{(2)}, I^{(3)}$ as a
  function of   strength of onsite anharmonicity $\lambda$. The inset shows
  the corresponding plots for inter-particle anharmonicty.}
\label{Jvsnonlin}
\end{figure}

We now consider the effect of introducing anharmonicity  in the Hamiltonian of
the loop. We consider a quartic onsite potential of the form  $V(x)=
\lambda x^4/4$ at all sites of the loop. We fix all other parameters to have
the same values as the pinned system studied earlier. We plot in Fig.~(\ref{Jvsnonlin})  the
currents $I^{(1)}, I^{(2)}, I^{(3)}$ in different regions  as a function of
the strength of anharmonicity $\lambda$.   
We see that CM decreases with anharmonicity.
Interestingly  we find that for some value of $\lambda$ the current on the
lower arm ($I^{(3)}$) becomes \emph{ exactly zero}.  Note also that the
current on the upper arm seems to vanish at some value of $\lambda$.  
The inset of Fig.~(\ref{Jvsnonlin}) shows the effect of an interparticle
anharmonic potential of the form $\lambda (x_l-x_{l+1})^4/4$ between particles
on the loop. We see a similar reduction of $CM$ as in the onsite case though
there are some qualitative differences.   
For macroscopic systems where the effective mean free path due
to anharmonicity becomes much smaller than the system size it is expected that
CM will not be observed. 
In Fig.~(\ref{Tprof}) we  show temperature profiles for a strongly
anharmonic case. In this  case, in contrast to the harmonic case, the
temperature  profile is {\emph{not}} non-monotonic and correspondingly there
is no CM.

{\emph{Conclusions}}:
Motivated by  studies of CM in quantum mesoscopic systems, in
this paper we have studied models of energy transmission in classical
oscillator chains. We have considered an oscillator loop connected to two external
leads. The system is made asymmetric by either making the two arm lengths of the
loop different or by introducing impurities.  
For single-frequency sound waves  we find that CM is obtained over particular frequency ranges. For the case where
the network is connected to heat baths which send waves at all frequencies we
find absence of CM. This is true for various 
parameter sets that we have tried,  but we do not have a  proof for this
result. We find that CM in the presence of heat baths can be
obtained if we introduce a pinning potential in the leads so that
only a narrow band of frequencies are allowed to pass through the loop. 
While we have reported results for a small loop, we have checked that for
harmonic systems CM is obtained for much larger sizes also.   
Finally we have looked at the effect of anharmonicity on
CM.  Our simulations show that CM is
reduced but not completely destroyed in the presence of small anharmonic
interactions. We also find the remarkable effect that with appropriate
choice of parameters, the current in one arm can be made to exactly vanish. 
We expect that CM in phononic heat  transport will be
observable experimentally in mesoscopic systems such as insulating nanotubes   
where the effective mean free path for inelastic phonon scattering is large
compared to system size.

RM acknowledges 
financial support from Grant MOSAICO (Spain) and thanks TIFR, Mumbai and RRI,
Bangalore for kind hospitality.  AMJ thanks  DST, India for financial support.


\end{document}